\documentclass[twocolumn,showpacs,preprintnumbers,amsmath,amssymb]{revtex4}
\usepackage{epsfig,amsfonts}
\usepackage{amsmath}
\usepackage{bbm}
\usepackage{graphicx,psfrag,rotating}
\usepackage{graphics}
\usepackage{txfonts}
\usepackage{graphicx}% Include figure files
\usepackage{dcolumn}% Align table columns on decimal point
\usepackage{bm}% bold math

\begin{document}

%\preprint{APS/123-QED}%

\title{ Optomechanical Entanglement in the Presence of Laser Phase Noise }

\author{R.Ghobadi$^1,^2$ A.R.Bahrampour$^2$ and C.Simon$^1$}
\affiliation{$^1$ Institute for Quantum Information Science
and Department of Physics and Astronomy, University of
Calgary, Calgary T2N 1N4, Alberta, Canada\\$^2$Department of Physics, Sharif University of Technology, Tehran, Iran}

\date{\today}

\begin{abstract}

We study the simplest optomechanical system in the presence of laser phase noise using the covariance matrix formalism. We show that the destructive effect of the phase noise is especially strong in the bistable regime. This explains why ground state cooling is still possible in the presence of phase noise, as it happens far away from the bistable regime. On the other hand, the optomechanical entanglement is strongly affected by phase noise.

\end{abstract}

\maketitle
\section{\label{intro}Introduction }

Observing quantum effects for macroscopic objects like mechanical oscillators, if not impossible\cite{Penrose00}, is extremely hard due to the interaction of the system
with its environment. According to this explanation if one properly isolates the system of interest from its environment it would be possible to observe bizarre quantum effects like nonlocality and entanglement at macroscopic level. One promising proposal to this end is to use an electromagnetic resonant  system coupled to a mechanical oscillator. This type of system has been realized in various contexts and scales, including optomechanical systems, electromechanical systems \cite{Teufel},  Cooper-pair boxes \cite{Tian04}, single electron transistors \cite{Naik06} and quantum dots \cite{Wilson04}. Using this technique recently the first man made device has been prepared in the ground state and its coherent interaction with a qubit has been observed \cite{OConnell10}.

It has already been shown\cite{Vitali07} that optomechanical systems can exhibit entanglement for small enough decoherence and strong enough optomechanical coupling. For typical optomechanical systems cryogenic cooling is not sufficient to reach the quantum regime. To further cool the mechanical resonator one can use side band laser cooling by a red detuned laser. The laser serves both to cool the mechanical oscillator and to enhance  the optomechanical coupling constant. In this way the strong coupling regime has been reached recently\cite{Groblacher09}. But the price to be paid is that a new source of noise is introduced in our system, namely laser phase noise (LPN).
 
The conclusion of early studies of the effect of LPN \cite{Diosi08}, based on modeling the LPN as white noise, suggested that with the current level of LPN ground state cooling is impossible. This is in contrast with several experiments in which low phonon numbers have been observed. This discrepancy has been addressed in \cite{Rabl09} by considering a more realistic model for the LPN. 

Here we consider the effect of realistic LPN on optomechanical entanglement. It turns out that optomechanical entanglement is very sensitive to this additional source of noise. It is particularly sensitive in the bistable regime, where the maximum entanglement is achieved in the absence of noise \cite{Genes2,Ghobadi11}.
 
The paper is organized as follows: Section II introduces the optomechanical system and establishes the notation. We obtain the governing equations in the presence of LPN in this section. In section III we introduce the LPN model. Section IV discusses the optomechanical cooling. We recover the well-known effect of LPN on the ground state cooling as a special case of our general results. Section V discusses the optomechanical entanglement in the presence of LPN. Section VI is a summary and conclusion.

\section{\label{ System} The System}

Our system is a  high finesse Fabry-Perot cavity with one tiny end mirror. This mirror can move under the influence of the radiation pressure inside the cavity and in the same time undergoes brownian motion as a result of its interaction with the environment. The system is driven by a laser with frequency $\omega_{L}$ with power $P$. The Hamiltonian of the system
is given by $H=H_{0}+H_{int}$ where
\begin{equation}
H_{0}=\hbar\omega_{c}a^{+}a+\frac{\hbar\omega_{m}}{2}(q^{2}+p^{2})
\end{equation}
\begin{equation}
H_{int}=\hbar G_{0}a^{+}aq+i\hbar E(a^{+}e^{-i\omega_{L}t}e^{-i\varphi(t)}-ae^{i\omega_{L}t}e^{i\varphi(t)})
\end{equation}
Eq.(1) describe the free Hamiltonian of the cavity mode and mechanical oscillator, respectively. $\omega_{c}$ and $a$ are frequency and annihilation operator
of the cavity mode,respectively, $\omega_{m}$, $q$,$p$ are frequency
and dimensionless position and momentum operator of the mirror,respectively.The first term in Eq.(2) describes the optomechanical interaction in which $G_{0}=\frac{\omega_{c}}{L}\sqrt{\frac{\hbar}{m\omega_{m}}}$
is the single photon coupling constant. The second term in Eq.(2) describes the cavity pumping with Laser which its phase variation in time is given by $\varphi(t)$ and $E=\sqrt{\frac{2P\kappa}{\hbar\omega_{L}}}$ where $P$, $\omega_{L}$ are the input laser power and frequency
respectively.

In the frame rotating with frequency $\omega_{L}t+\varphi(t)$
the equations of motion in the presence of damping and noise are
\begin{equation}
\dot{q}=\omega_{m}p\end{equation}
\begin{equation}
\dot{p}=-\omega_{m}q-\gamma_{m}p+G_{0}a^{+}a+\xi(t)\end{equation}
\begin{equation}
\dot{a}=-(\kappa+i\Delta_{0})a+ia\dot{\varphi}+iG_{0}aq+E+\sqrt{2\kappa}a_{in}\end{equation}
where $\Delta_{0}=\omega_{c}-\omega_{L}$ and $a_{in}$ is the vacuum input noise.
The mechanical and the optical input noise operators are fully characterized by their correlation function which in the Markovian approximation given by
\begin{equation}
\langle a_{in}(t)a_{in}^{+}(t')\rangle=\delta(t-t').\end{equation}
\begin{equation}
\frac{\langle\xi(t)\xi(t')+\xi(t')\xi(t)\rangle}{2}=\gamma_{m}(2\bar{n}+1)\delta(t-t').\end{equation}
where $\overline{n}=[exp(\frac{\hbar\omega_{m}}{k_{B}T})-1]^{-1}$
is the mean thermal phonon number and $k_{B}$ is Boltzmann's constant.
The nonlinear Eqs. (3,4) can be linearized by expanding the operators around their steady state values $\hat{O_{i}}=\bar{O}_{i,s}+\delta O_{i}(t)$ ,where $\hat{O}_{i}=q\; ,p ,a $.

The steady state solution of the system is given by $q_{s}=\frac{G_{0}\mid\alpha_{s}\mid^{2}}{\omega_{m}}$, $p_{s}=0$,$\alpha_{s}=\frac{E}{\kappa+i(\Delta_{0}-G_{0}q_{s})}$,
 where $\alpha_{s}$,$q_{s}$,$p_{s}$ are the values for cavity amplitude, position and momentum of mechanical oscillator,respectively. The last of these relations is a third order polynomial equation for $\alpha_s$, which can give rise to optomechanical bistability\cite{Dorsel83,Karuza} . As can be seen from Fig. 1 for strong enough input power the intracavity power admit two solutions. Fig. 1 shows the hysteresis loop for the intracavity power. Consider $P_{cav}$ initially on the lower stable branch (I in Fig. 1). By increasing the input power we approach the end of the first stable branch.
Increasing the input power beyond the end of the stable branch causes the power inside the cavity to switch to the second stable branch (II in Fig. 1). For input power larger than this value the cavity power is given by the upper branch. If the input power is decreased again, the cavity power follows a hysteresis loop. The parameter which describes the distance from the end of the stable branches is called bistability parameter which is defined by\cite{Genes08,Ghobadi11}
\begin{equation}
\eta=1-\frac{G^{2}\Delta}{\omega_{m}(\kappa^{2}+\Delta^{2})}
\end{equation}
which is a real number between zero and one as long as the system has a stationary state\cite{Vitali07}. As can be seen from Fig.1, $\eta$ decreases when approaching the bistable regime and becomes equal to zero at the end of each stable branch.
\begin{figure}
\scalebox{0.5}{\includegraphics*[viewport=0 0 650 300]{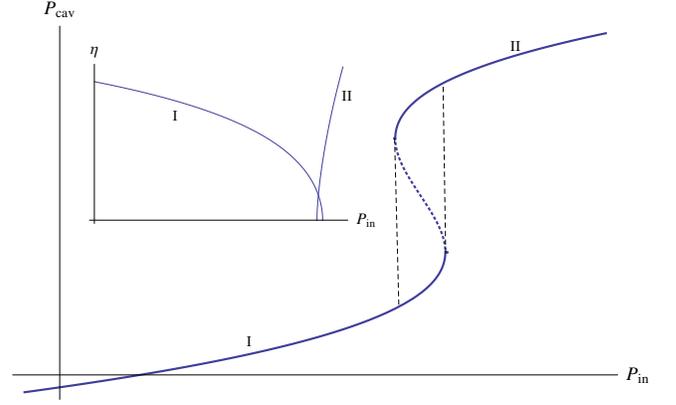}}
\caption{\label{fig1}  Bistability of the intracavity power with respect to the input power\cite{Ghobadi11}. The solid and dotted lines correspond to the stable and unstable branches respectively. The inset shows the bistability parameter $\eta$ for the two stable branches. The end of each stable branch corresponds to $\eta= 0$.
}
\end{figure}

Introducing amplitude and phase quadratures as $X=\frac{\delta a+\delta a^{+}}{\sqrt{2}}$ and $Y=\frac{\delta a-\delta a^{+}}{\sqrt{2}i}$ and corresponding noises $X_{in}$and $Y_{in}$ we have

 \begin{equation}
\delta\dot{q}=\omega_{m}\delta p\end{equation}

 \begin{equation}
\delta\dot{p}=-\omega_{m}\delta q-\gamma_{m}\delta p+G\delta X+\xi(t)\end{equation}

 \begin{equation}
\delta\dot{X}=-\kappa\delta X+\Delta\delta Y+\sqrt{2\kappa}X_{in}\end{equation}

 \begin{equation}
\delta\dot{Y}=-\kappa\delta Y-\Delta\delta X+G\delta q+\sqrt{2\kappa}Y_{in}+\sqrt{2}\alpha_{s}\dot{\varphi}\end{equation}
where $G=\sqrt{2}G_{0}\alpha_{s}$ , $\Delta=\Delta_{0}-G_{0}q_{s}$
are the enhanced optomechanical coupling rate and effective detuning.
Introducing $u^{T}(t)=(q(t),p(t),X(t),Y(t))$ and $n^{T}=(0,\xi(t),\sqrt{2\kappa}X_{in},\sqrt{2\kappa}Y_{in}+\sqrt{2}\alpha_{s}\dot{\varphi})$,  Eq.(8-11) can be written in a compact form

\begin{equation}
\dot{u}(t)=Au(t)+n(t)\end{equation}
 where
\begin{equation}
A=\left(\begin{array}{cccc}
0 & \omega_{m} & 0 & 0\\
-\omega_{m} & -\gamma_{m} & G & 0\\
0 & 0 & -\kappa & \Delta\\
G & 0 & -\Delta & -\kappa\end{array}\right)\end{equation}

The solution of Eq.(13) is given by $u(t)=M(t)u(0)+\int_{0}^{t}dsM(s)n(t-s)$ where $M(t)=exp(At)$. The system reaches steady state if real part of the eigenvalues of $A$ be negative which is the case if
$0<\eta<1$\cite{Vitali07}.

Since the initial state of the system is Gaussian and the dynamical
equation of system is linear in creation and annihilation operator
both for cavity and mechanical mode the state of the system remain
Gaussian at all time. A Gaussian state is fully characterized by its
covariance matrix which is defined at any given time $t$ by $V_{ij}(t)=\frac{\langle u_{i}(t)u_{j}(t)+u_{j}(t)u_{i}(t)\rangle}{2}$. Using the solution of Eq.(13) one obtain the following equation for steady state covariance matrix
\begin{equation}
AV+VA^{T}=-D.\end{equation}
where the diffussion matrix is given by

\begin{equation}
D_{ij}=\frac{1}{2}\int_{0}^{\infty}ds(M_{ik}(s)\langle\{n_{k}(s),n_{j}(0)\}\rangle+M_{jk}(s)\langle\{n_{i}(0),n_{k}(s)\}\rangle).
\end{equation}
Using Eq.(6,7), we get
$D=Diag[0,\gamma_{m}(2\bar{n}+1),\kappa,\kappa+N]$
and $N=\int_{0}^{\infty}dsM_{44}(s)\langle\dot{\varphi}(s)\dot{\varphi}(0)\rangle$ describes the contribution of the LPN to the decoherence.
Using Eq.(15) with the modified diffusion matrix the effect of arbitrary LPN on the behaviour of the optomechanical system can be obtained.

\section{\label{ Noise} The Noise Model}
The dynamics of phase for an ideal single-mode laser far above the threshold is given by \cite{Haken,Kennedy86}
\begin{equation}
\ddot{\varphi}+\gamma_{c}\dot{\varphi}=\xi_{\varphi}(t)\end{equation}
 in which $\xi_{\varphi}(t)$ is a Gaussian random variable obeying\cite{Haken}
\begin{equation} \langle\xi_{\varphi}(t)\xi_{\varphi}(t^{'})\rangle=2\gamma_{c}^{2}\Gamma_{L}\delta(t-t^{'}).\end{equation}
Here $\Gamma_{L}$ and $\gamma_{c}^{-1}$ describe the laser linewidth and finite correlation time of phase noise. Using Eqs. (7) and (18), the phase noise spectrum is given by \begin{equation}
S_{\dot{\varphi}}(\omega)=\frac{2\Gamma_{L}}{1+\frac{\omega^{2}}{\gamma_{c}^{2}}}\end{equation}
Note that for $\gamma_{c}\rightarrow\infty$ one recovers the white noise, while for finite correlation time and $\omega\gtrsim\gamma_{c}$ the power spectrum of LPN is greatly reduced. This means that a white noise model may greatly overestimate the effects of LPN.
For this specific noise model we obtain
 \begin{equation}
N=2\alpha_{s}^{2}\gamma_{c}\Gamma_{L}\int_{0}^{\infty}dsM_{44}(s)e^{-\gamma_{c}s}.\end{equation}
The general expression for $N$ is very complicated and will not be included here. But it is still possible to get some insight about the behaviour of $N$ qualitatively. $M_{44}(s)$ is one of the elements of the matrix $M(s)=\exp(As)$. Far from the bistable regime, all eigenvalues of $A$ are significantly smaller than zero, and all elements of $M(s)$ decay with $s$, effectively limiting the range of the integral in Eq. (20). However, as one approaches the bistable regime, one of the eigenvalues of $A$ approaches zero, and the range of the integral increases until it is limited only by the exponential $e^{-\gamma_c s}$. One thus expects $N$ to take its maximum values in the bistable regime.

%To do so we note that we can write $M_{44}(s)=\sum_{i=0}^{4}e^{x_{i}s}f(x_{i})$
% where $x_{i}$ are the eigenvalues of $A$ and $f(x)$ is a rational function. using this expression in Eq.(20) one finds $N=2\alpha_{s}^{2}\gamma_{c}\Gamma_{L}\sum_{i=1}^{4}\frac{f(x_{i})}{\gamma_{c}-x_{i}}$.
%  As we come close to the bistable region one of the eigenvalues approaches zero while other one are still very large. This means that the $N$ takes its maximum at close to the bistability region.

 Fig.2 shows $N/\kappa$ as a function of effective detuning. It can be seen from Eq.(20) and also in Fig. 2 that the noise is proportional to laser bandwidth, i.e $\Gamma_{L}$. One also sees that $N/\kappa$ is maximum for $\eta\thicksim0$, in agreement with the above argument. We would like to emphasize that the above argument is not restricted to the specific LPN model which we use here. In fact as long as the
 system admits a stationary solution (this guarantees that $M_{44}(s)$ is a monotonic decreasing function of $s$), the above argument is valid for any LPN model with a finite correlation time.

\begin{figure}
\scalebox{0.45}{\includegraphics*[viewport=0 0 650 400]{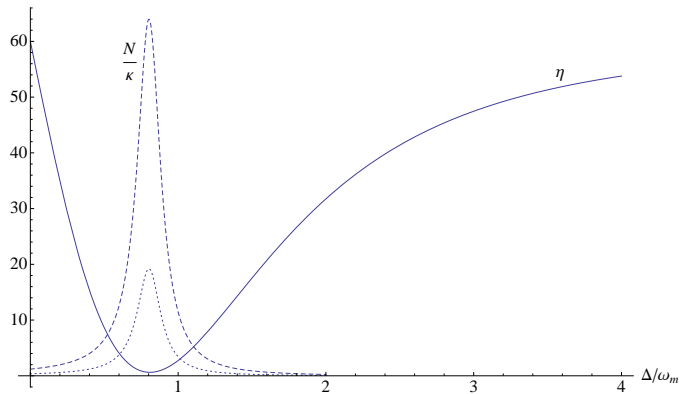}}
\caption{\label{fig1} Plot of $N/\kappa$ and $\eta$ as a function of effective detuning. We consider a Fabry-Perot cavity with length $L=1mm$ and $\kappa=1.4\omega_{m}$, driven by a laser with $\lambda=810 nm$ and input power $50 mW$. The mechanical oscillator frequency, damping rate and mass are $10$ MHz,$100$ Hz and $5$ ng respectively. The solid line corresponds to $\eta$ (multiplied by 60). The dashed and dotted dashed curves correspond to $\Gamma_{L}=100$ Hz and 30 Hz respectively.}
\end{figure}

\section{\label{cooling} Optomechanical Cooling}
For completeness we study the optomechanical cooling in the presence of LPN.
Solving Eq. (15) one can obtain the phonon number $\bar{n}_{m}=\frac{ V_{11}+V_{22}-1}{2}$. The quantum limit of phonon number can be derived by assuming a high mechanical quality factor and low temperature environment,i.e. $\frac{\omega_{m}}{\gamma_{m}}>>1$
and $\frac{\kappa}{\bar{n}\gamma_{m}}>>1$. For $\eta\sim1$,$\kappa<<\omega_{m}$ and $\Delta=\omega_{m}$ we find
\begin{equation}
\bar{n}_{m}=\frac{\kappa^{2}}{4\omega_{m}^{2}}+\frac{N}{4\kappa}
\end{equation}

\begin{equation}
N=\frac{2\alpha_{s}^{2}\Gamma_{L}\gamma_{c}(\gamma_{c}+\kappa)}{\omega_{m}^{2}+(\gamma_{c}+\kappa)^{2}}\end{equation}

Assuming $\omega_{m}>>\gamma_{c}>>\kappa$ we find
\begin{equation}
\bar{n}_{m}=\frac{\kappa^{2}}{4\omega_{m}^{2}}+\frac{\alpha_{s}^{2}}{2\kappa}S_{\dot{\varphi}}(\omega_{m})\end{equation}
which is identical to Eq. (23) in \cite{Rabl09} if one add the effect of residual thermal occupation. From Eq.(19,23) its clear that as long as $\omega_{m}>>\gamma_{c}$ the LPN effect on ground state cooling can be small. For $\gamma_{c}<<\kappa$ the second term in Eq.(23) becomes  $\frac{\alpha_{s}^{2}}{2\gamma_{c}}S_{\dot{\varphi}}(\omega_{m})$. Intuitively the possibility of ground state cooling in the presence of LPN can be understood as a result of the fact that ground state cooling happen around $\eta\sim1$, far from the bistability regime, and thus in a region where the effect of LPN is small. See also the discussion of ground state cooling from the point of view of the bistability parameter in Ref. \cite{Ghobadi11}.

\begin{figure}[h!]

\epsfig{file=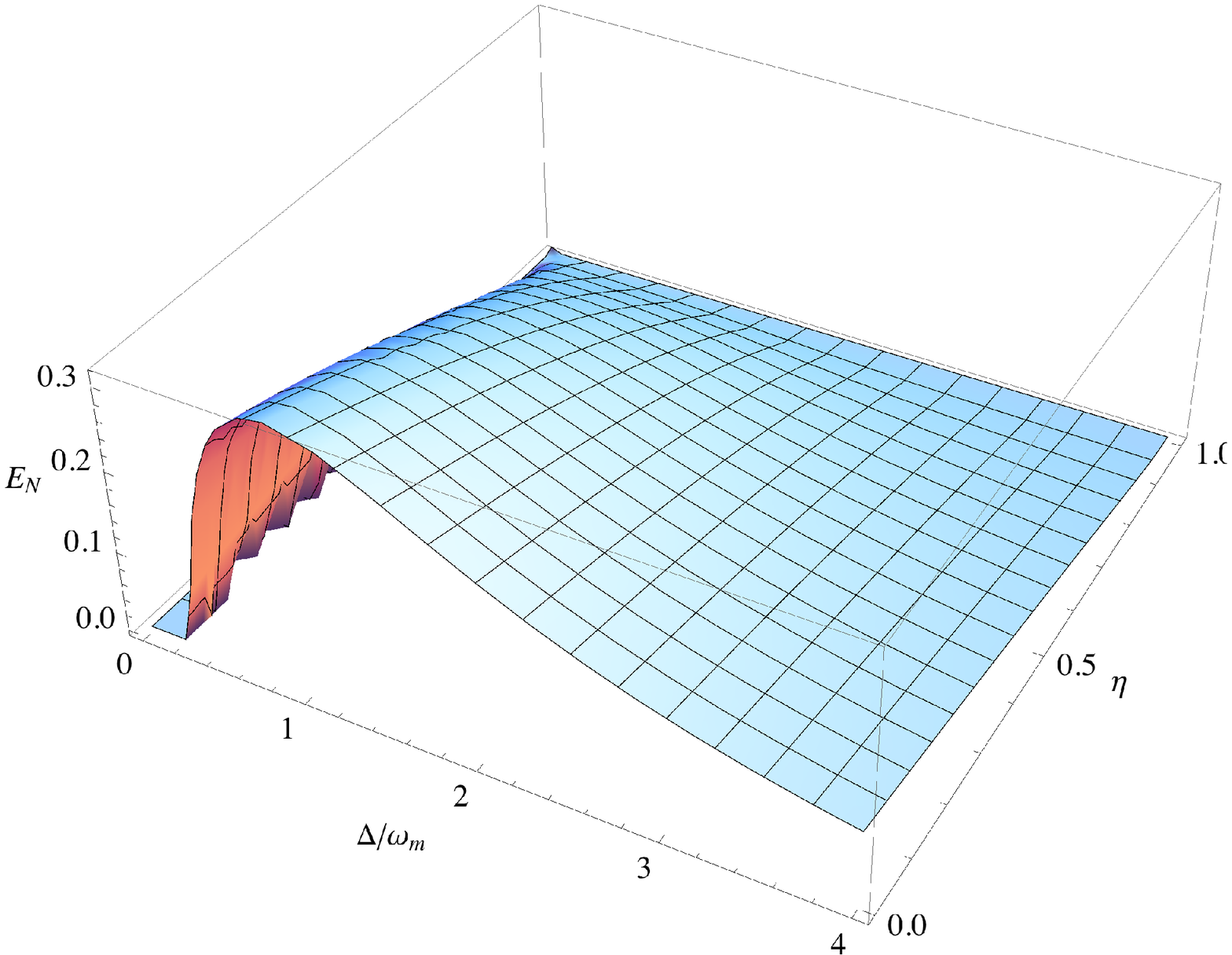,width=0.9\columnwidth}
\epsfig{file=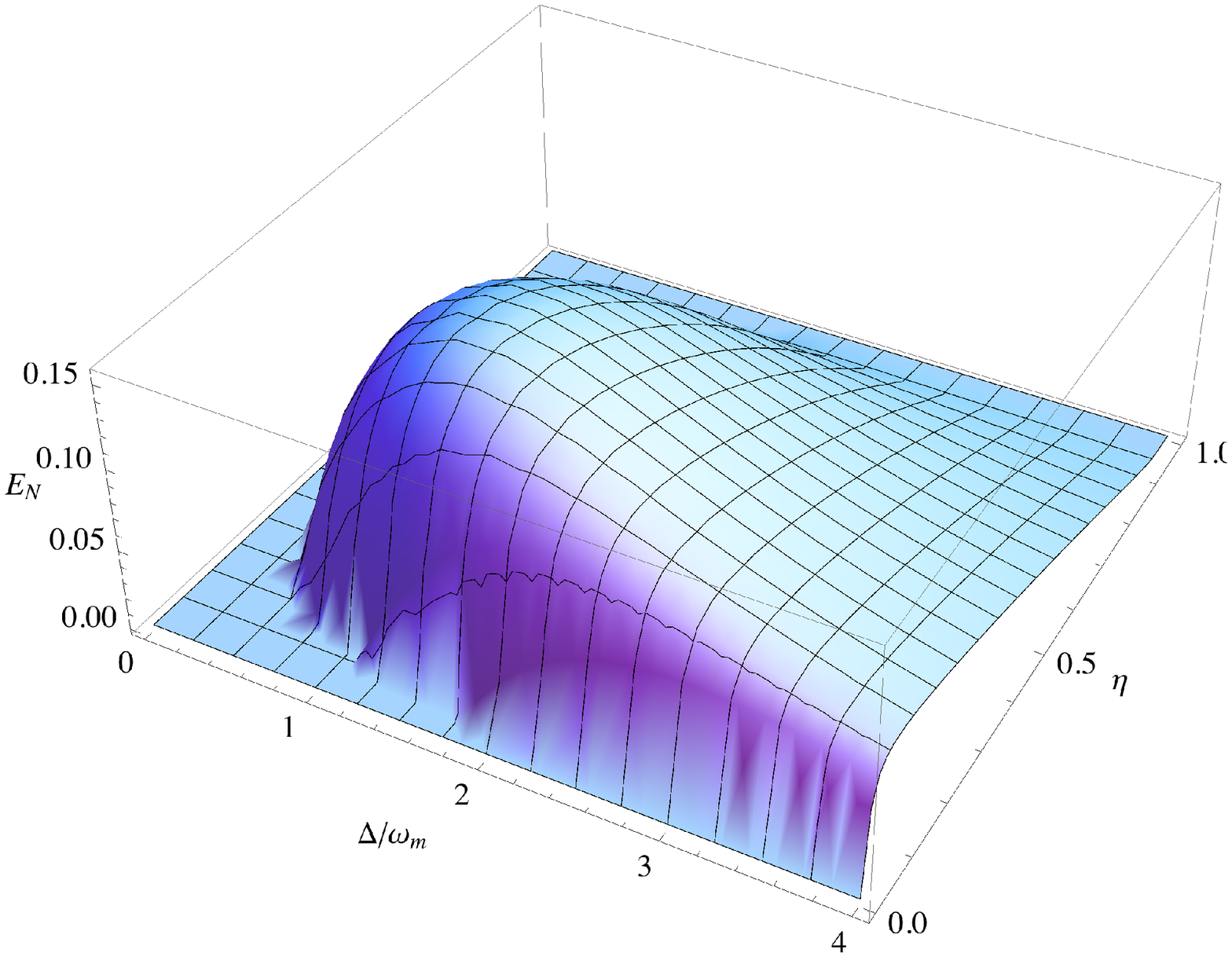,width=0.9\columnwidth}
\epsfig{file=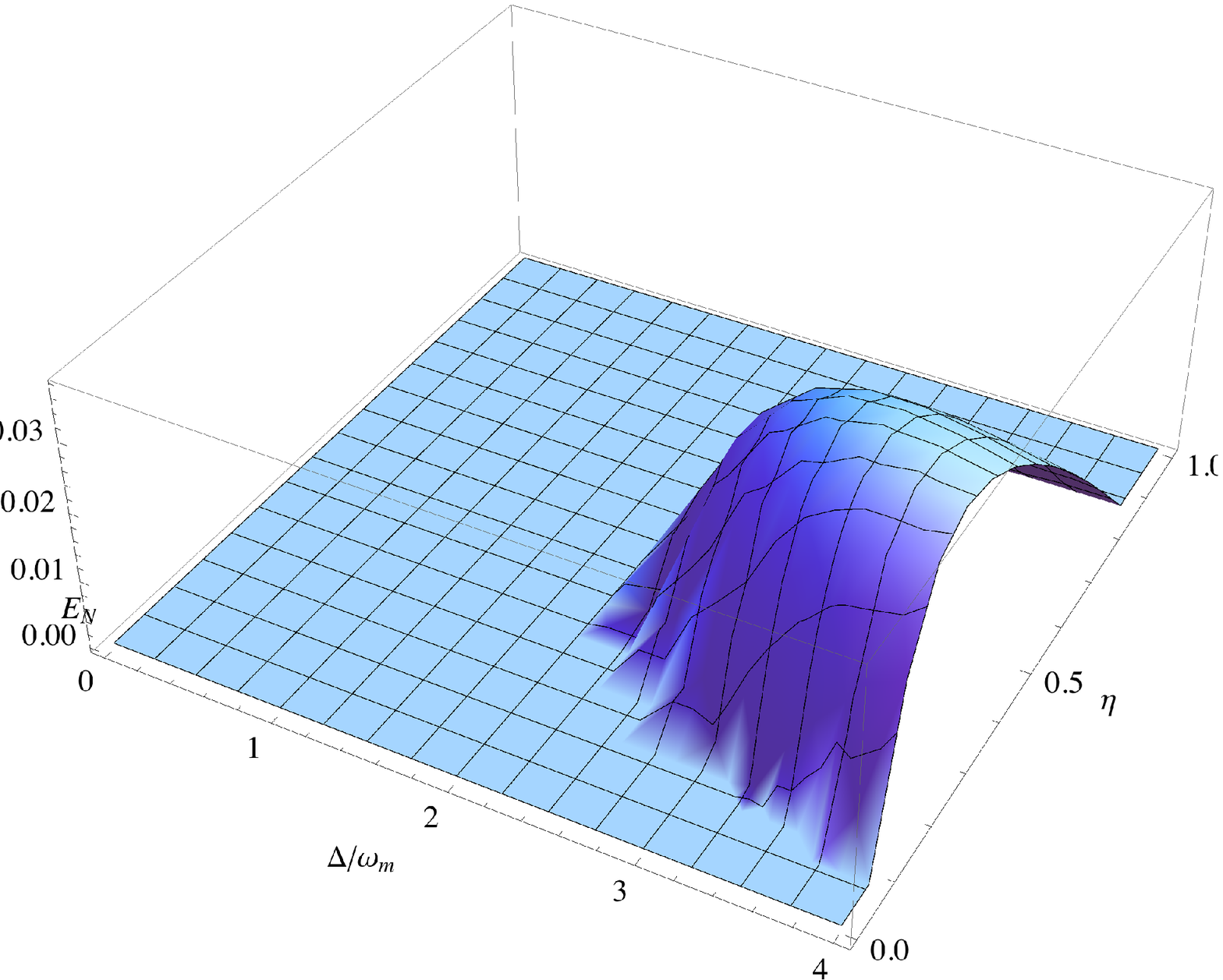,width=0.9\columnwidth}

\caption{\label{fig1} Entanglement as a function of the bistability parameter $\eta$ and the normalized detuning $\Delta/\omega_m$ for three different values of the laser linewidth $\Gamma_L$, from top to bottom: $\Gamma_L=0$, 10 Hz and 100 Hz. The other parameters are the same as for Fig. 2.}
\end{figure}

\section{\label{Entanglement} Optomechanical Entanglement}

We use logarithmic negativity as the measure of entanglement, which is defined as\cite{Adesso04}
\begin{equation}
E_{N}=max\{0,-ln(2\nu_{min})\}.\end{equation}
where $\nu_{min}$ is the smallest symplectic eigenvalue of the partially
transposed covariance matrix given by $\nu_{min}=\sqrt{\frac{\Sigma-\sqrt{\Sigma^{2}-4detV}}{2}}$, where $\Sigma=detA+detB-2detC$, and we represent the covariance matrix in terms of
\begin{equation}
V=\left(\begin{array}{cc}
A & C\\
C^{T} & B\end{array}\right).\end{equation}

Fig. 3 shows the entanglement as a function of the bistability parameter and the detuning for three different values of the laser linewidth. In the absence of phase noise, entanglement is maximal in the bistable regime, i.e. for $\eta \sim 0$. However, this changes dramatically as soon as the laser linewidth is non-zero. The entanglement goes to zero for $\eta=0$ for a linewidth as small as $\Gamma_L=10$ Hz. For $\Gamma_L=100$ Hz only a small amount of entanglement survives, and the region of maximum entanglement is comparatively far from the line $\eta=0$. It is noticeable that for the $\Gamma_L=100$ Hz entanglement survives only for $\Delta/\omega_m > 2$. This can be understood from Fig. 2, because the noise term $N$ becomes very small for these values of the detuning.

\section{\label{Summary and Conclusion} Summary and Conclusion}

We studied a generic optomechanical system in the realistic situation where the input laser bandwidth can not be ignored. We found that the LPN contribution to the decoherence, characterized by $N$, is particularly significant in the bistable regime ($\eta \sim 0$), and significantly suppressed elsewhere. This explains why ground state cooling is still possible, as it happens for $\eta\sim1$. In contrast, optomechanical  entanglement in the absence of LPN is maximal in the bistable regime. As a consequence, both the optimum region for the observation of entanglement and the amount of entanglement that can be achieved are strongly affected by LPN.

{\it Note added.} When this work was completed, we became aware of a recent related paper\cite{Abdi11}. In comparison, the authors of that paper use a different noise model, and they treat the laser phase and amplitude as additional dynamical variables. In contrast we keep the number of dynamical variables fixed and treat the LPN via an additional term in the diffusion matrix.

{\it Acknowledgments.} This work was supported by AITF and NSERC.

\end{document}